\documentclass[a4paper,12pt,english,german]{article}

\usepackage[pdftex]{graphicx}

\begin{document}

\centerline{\bf   Zero Discord  for Markovian Bipartite Systems }

\bigskip

M. Arsenijevi\' c$^{\dag}$\footnote{Email: fajnman@gmail.com}, J.
Jekni\' c-Dugi\' c$^{\ast}$, M. Dugi\' c$^{\dag}$,

\smallskip

$^{\dag}${Department of Physics, Faculty of Science, Kragujevac,
Serbia}

$^{\ast}${Department of Physics, Faculty of Science, Ni\v s,
Serbia}

\bigskip

{\bf Abstract} Recent observation that almost all quantum states
bear non-classical correlations [A. Ferraro et al, Phys. Rev. A
81, 052328 (2010)] may seem to imply that the Markovian bipartite
systems are practically deprived of zero discord states.
Nevertheless, complementary to the result of Ferraro et al, we
construct a model of a Markovian bipartite system providing zero
discord for arbitrary long time interval, that we term 'Markovian
classicality'. Our model represents a matter-of-principle formal
proof, i.e. a sufficient condition for the, otherwise not obvious,
existence of Markovian classicality. Interestingly enough, we are
not able to offer any alternative to the model. Physical relevance
of the model is twofold. First, the model is in intimate relation
to the topics of quantum information locality, quantum discord
saturation and quantum decorrelation. Second, the model is of the
general physical interest. It  pertains to a specific structure
(decomposition into parts/subsystems) of a composite system, not
to a special physical kind of  composite systems. Being a
characteristic of a structure, by definition, the model of
Markovian classicality is not a model of sudden death of discord.
We emphasize wide-range implications of our results.

\bigskip


\bigskip

{\bf I. INTRODUCTION}

\bigskip

"Quantum discord" is a common term for different measures of
non-classical correlations in composite (e.g. bipartite) quantum
systems [1-6]. Historically the first and probably the best known
is the so-called "one-way" discord (to be defined in Section II)
[1, 2]. The closely related "two-way" discord is even a more
stringent criterion for classical correlations. The only states of
a bipartite system providing zero two-way discord are the
so-called classical-classical (CC) states (see Definition 1,
Section II).

A recent analysis of the one-way-discord dynamics provides a
remarkable observation [7]. The authors find [7]: "that for almost
all states of positive discord, the interaction with any
(non-necessarily local) Markovian bath can never lead either to a
sudden, permanent vanishing of discord, nor to one lasting a
finite time-interval". In effect, not only  sudden death of
discord cannot be expected, but Markovian dynamics only leads us
{\it asymptotically close} to a zero-discord state. From this
result one may possibly expect the Markovian bipartite systems are
practically deprived of zero discord states.

However, the analysis in [7] does not rule out that there can be
zero discord {\it for all times}. If a state starts with zero
discord, it could be zero discord for all times [8]. Thus,
complementary to Ferraro et al [7], Markovian dynamics may
probably provide {\it non-asymptotic} zero-discord  for a
bipartite system in a {\it long} time interval ('{\it Markovian
classicality}').

In this paper, our task is twofold. {\it First}, we are interested
in answering the following questions: Which kind, if any, of the
zero-discord states can provide Markovian classicality? Given an
answer (or a guess) to the first question: is there  a physical
model that can justify such zero-discord states dynamics? What are
the physical characteristics of such model(s)? {\it Second}, we
are interested in linking such model(s) to the {\it realistic}
physical systems and situations. While the first task bases itself
on the existing knowledge about the discord dynamics, the second
one welcomes a change in perspective to the composite quantum
systems.

Regarding the first task: As two-way discord tends to be larger
than one-way discord [9], we consider the zero {\it
two}-way-discord (the CC) states. Accordingly, Markovian
classicality is defined by zero-discord as a 'constant of motion',
that is by  the open system's dynamics as a dynamical map from one
to another CC state (cf. Definition 1, Section II). Then we {\it
construct}, not deduce, such a model; the model satisfies both the
$C$- and the $P$-criterion for classicality [10]. Our approach is
a formal mathematical analysis that leads us to the simplest
possible model of tensor-product state for the open system.
Interestingly enough, {\it we are not able to find any alternative
to the model}. The model reveals a number of physically
interesting observations
 such as relations to the quantum information
locality [11-13], quantum discord saturation [14] and quantum
decorrelation [15, 16] topics (Section III.A).

Regarding the second task, we emphasize importance of "structure"
(decomposition into parts/subsystems) of a composite system. We
show (Section III.B) the model of Markovian classicality is a
matter of a special structure of a composite system. The composite
systems not describable by such structure may be deprived of zero
discord states. So, being a matter of structure,  the model is of
the general physical relevance. The following example illustrates
this is implicit in the foundations of the quantum information
science. Consider a three-qubit system, $\mathcal{C} = 1+2+3$, and
its bipartite structures, $1+S_1$ and $S_2+3$, where the bipartite
systems $S_1=2+3$ and $S_2=1+2$.  As it is well known from quantum
teleportation [17], the $\mathcal{C}$'s state $\vert \phi\rangle_1
\vert \Phi^{+}\rangle_{S_1}$, where $\vert \Phi^{+}\rangle_{S_1} =
(\vert 0\rangle_2 \vert 0\rangle_3 + \vert 1\rangle_2 \vert
1\rangle_3)/2^{-1/2}$, can be re-written as $\sum_i \vert
\chi_i\rangle_{S_2}\vert i\rangle_3/2$, where the $S_2$'s states
represent the Bell states [18] for the pair $1+2$. The point is
that for the $1+S_1$  structure, the state is tensor-product and
therefore not bearing any correlations between the $1$ and $S_1$
systems, while there is entanglement in the $S_2+3$ structure. So,
for the closed $\mathcal{C}$ system, the structure $1+S_1$ bears
Markovian classicality, which is not the case for the $S_2+3$
structure. Being a characteristic of a structure, the Markovian
classicality model is not a model of sudden death of discord.

This paper is organized as follows. In Sec. II, we give a precise
formulation of the task and design the model supporting Markovian
classicality. Information theoretic  analysis of the model in Sec.
III gives rise to a need to relax the definition of classicality.
In Sec. IV, we introduce  approximate classicality and recognize a
model implementing such approximate classicality. Section V is
discussion where we emphasize wide-range implications of our
results. Section VI is Conclusion.

\bigskip

{\bf II. THE MODEL}

\bigskip

One-way quantum discord for the $S+S'$ system,
$D^{\leftarrow}(S|S') = I(S:S') - J^{\leftarrow}(S|S') \ge 0$, and
von Neumann entropy of a state $\rho$, $\mathcal{S} = - tr
\rho\ln\rho$. Both the total mutual information, $I(S:S') =
\mathcal{S}(S) + \mathcal{S}(S') - \mathcal{S}(S,S')$, and the
classical correlations, $J^{\leftarrow} (S\vert S') =
\mathcal{S}(S) - \inf_{\{\Pi_{S'i}\}}\sum_i \vert c_i \vert^2
\mathcal{S}(\rho_S\vert_{\Pi_{S'i}})$--where
$\rho_S\vert_{\Pi_{S'i}} = I_S \otimes \Pi_{S'i} \rho I_S \otimes
\Pi_{S'i}$ is the state remaining after a selective quantum
measurement defined by the projectors $\Pi_{S'i}$--are
non-negative. The CC states are the only states fulfilling the
condition $D^{\leftarrow}(S|S') = 0 = D^{\rightarrow}(S|S')$.

\smallskip

\noindent {\it Definition 1.} An open quantum system, $C$,
consisting of two subsystems, $S$ and $S'$, is said to bear {\it
Markovian classicality} if and only if it can be described by a
classical-classical (CC) state in {\it  long} time-interval. A CC
state is of the form $\sum_{m,n} \omega_{mn} P_{Sm} \otimes
\Pi_{S'n}$, where the real numbers $\omega_{mn} \ge 0$ and
$\sum_{m,n} \omega_{mn} tr_S P_{Sm} tr_{S'} {\Pi}_{S'n} = 1$ for
the projectors $P_{Sm}$ and $\Pi_{S'n}$ on the respective Hilbert
spaces.

\noindent For separable $\omega_{mn} = p_m q_n, \forall{m,n}$,
such that $\sum_m p_m tr_S P_{Sm} = 1 = \sum_n q_n tr_{S'}
\Pi_{S'n}$, one obtains the tensor-product states, $\rho_S \otimes
\rho_{S'}$, as a special kind of CC states. Physically, the
composite system $C$ may be e.g. a pair "object of measurement +
apparatus" or "the internal + the center-of-mass" degrees of
freedom of the Brownian particle [19] (and the references
therein).

As typical of open systems, we assume a coarse-grained time scale
for the open system's dynamics [19]. On the other hand, the time
scale characteristic for {\it Markovian} dynamics we are {\it
exclusively interested in} is bounded also from the above
[19]--zero discord is not required for arbitrary long
time-interval either. This way understood classicality does both:
permits non-classicality for the time intervals shorter than e.g.
the "decoherence time", $\tau_D$, for the open system, $C$, and
still assumes the long time intervals for the possible thermal
relaxation of the open system, as well as for the "recurrence
time" regarding the closed system, $C+E$, where $E$ is the $C$'s
environment.

Definition 1 directly sets the following constraint on
constructing a Markovian classicality model:

\smallskip

\noindent Classicality Constraint: {\it Two-way quantum discord is
exactly zero in every instant in time before eventual
thermalization of the open system.}

\smallskip

Getting into details, we detect the following obstacles to
construct a model fulfilling the Classicality Constraint. First,
initial non-zero discord in $S+S'$ system; Second, interaction
between $S$ and $S'$; Third, the common environment, $E$, for $S$
and $S'$; Fourth, non-completely positive dynamics for the $S'$
system; Fifth, the initial non-tensor-product state for $C$ and
$E$; Sixth, arbitrary initial zero-discord state for $C$.

 The origin of these obstacles  is respectively
as follows: First, an initial non-zero discord state cannot fulfil
the classicality condition. e.g. The dynamic transition

\begin{equation}
\sum_i \lambda_i \rho_{Si} \otimes \rho_{S'i} \to \sum_{m,n}
\omega_{mn} \vert m\rangle_S\langle m \vert \otimes \vert
n\rangle_{S'}\langle n \vert
\end{equation}

\noindent is not allowed as long as the rhs of Eq. (1) refers to a
continuous time interval [7]. There are at  least three
 ways for dynamically obtaining a non-zero-discord state: Interaction
between $S$ and $S'$, the common environment for $S$ and $S'$, and
 non-completely positive dynamics for the open system $S'$. Markovian
dynamics requires the tensor product initial state $\rho_C \otimes
\rho_E$ [19]. Finally, in general, the external (e.g.
experimentally uncontrollable) local influence can raise the
initially zero discord [3, 20-22]. The local operations exerted on
$S$ and/or on $S'$, the rhs of Eq. (1), can give rise to
non-zero-discord final state. The {\it only state immune} to this
(yet for the  completely positive dynamics) is actually the
tensor-product state, $\rho_S \otimes \rho_{S'}$.

Bearing all this in mind, the {\it only} option we offer is the
following model:

\begin{equation}
S + (S'+E)
\end{equation}

\noindent where the subsystem $S$ does not interact with any other
subsystem ($S'$ and $E$) while assuming Markovian and completely
positive dynamics for the open system, $S'$, and the
tensor-product initial state $\rho_S \otimes \rho_{S'} \otimes
\rho_E$ for the total system, see Fig.1. In principle, both $S$
and $S'$ can be composite systems themselves.

\begin{figure}
\includegraphics[width=0.5\textwidth]{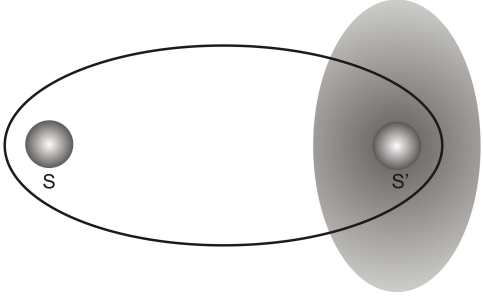}\\
\caption{Schematic illustration of the model Eq. (2). The
composite system $C=S+S'$ is distinguished by the elliptic line.
The gray area designates the environment $E$ in interaction with
$S'$. The $S$ system does not interact with both $S'$ and $E$.
Physically, the $S$ and $S'$ systems can represent respectively
e.g. the "relative (internal)"- and the center-of-mass-degrees of
freedom of the Brownian particle $C$. The pair $S+S'$ is described
by Eq. (5) and by the zero two-way discord, $D^{\leftarrow}(S\vert
S') = 0 = D^{\leftarrow}(S'\vert S)$, in every instant in
time.}\label{Fig. 1}
\end{figure}

Formally, the model Eq. (2) is defined by the Hilbert state space
for the total system $\mathcal{H} = \mathcal{H}_S \otimes
\mathcal{H}_{S'} \otimes \mathcal{H}_E$ and by the Hamiltonian of
the total system:

\begin{equation}
H = H_S + H_{S'} + H_E + H_{S'E}
\end{equation}

\noindent where the last term on the rhs of Eq. (3) represents
interaction between $S'$ and $E$. Then the unitary operator for
the total system separates as:

\begin{equation}
U(t) = U_S(t) \otimes U_{S'+E}(t) = \exp\{-\imath t H_S/\hbar\}
\otimes \exp[-\imath t (H_{S'} + H_E + H_{S'E})/\hbar],
\end{equation}

\noindent and provides unitary (the Schr\" odinger) dynamics for
both the $S$ system as well as for the  $S'+E$ system. Markovian
and completely positive dynamics of $S'$ does not introduce any
additional correlation for $S$ and $S'$. Then for the model Eq.
(2), one can write for the open system's state:

\begin{equation}
\rho_S(t) \otimes \rho_{S'}(t)
\end{equation}

\noindent {\it in every instant in time}, where $\rho_S(t) =
U_S(t) \rho_S(0) U_S^{\dag}(t)$ and $\rho_{S'}(t)$ is a solution
to a Markovian-type master equation. The proof of Eq. (5)
obviously follows from Eq. (4).

 From Eq.
(5) it easily follows: $\mathcal{S}(S,S') = \mathcal{S}(S) +
\mathcal{S}(S')$ and therefore the equalities
$D^{\leftarrow}(S|S') = 0 = D^{\rightarrow}(S|S')$ in every
instant in time. So, we can say {\it we have designed a model that
fulfills the very tight conditions for non-asymptotic zero-discord
classicality of a Markovian bipartite system}, Definition 1: (1)
the model Eq.(2)-(5) is distinguished, and (2) the open system's
dynamics is a completely positive map.

\bigskip

{\bf III. ANALYSIS OF THE MODEL}

\bigskip

The model Eqs. (2)-(5) is  {\it designed} so as to fulfill  the
Classicality Constraint, Section II. For the tensor-product
initial state, $\rho_S \otimes \rho_{S'}$, the subsystems $S$ and
$S'$ remain mutually exactly uncorrelated in every instant in
time, Eq. (5). In terms of [7]: the composite system's state
remains in the $\Omega_{\circ}$ set of zero-discord states, all
the time. As $[\rho_S \otimes I_{S'}, \rho_{S+S'}] = 0 = [I_S
\otimes \rho_{S'}, \rho_{S+S'}], \forall{t}$, the state
$\rho_{S+S'}$ Eq. (5) is a "doubly" lazy state [23]. Thus, we
point out a 'niche' for the bipartite system's Markovian
classicality. Most of the proofs recognized easy throughout the
remainder of this article are direct corollaries of the material
provided as the Supplemental Material.

\bigskip

 {\bf A. Quantum Information Locality and Classicality}

\bigskip

If one assumes the pure initial states for both $S'$ and $E$, then
Eqs. (2)-(4) directly give for the total system's instantaneous
state:

\begin{equation}
\rho_S \otimes \vert \Psi\rangle_{S'+E}\langle \Psi \vert
\end{equation}

\noindent and {\it vice versa}--given the above assumptions, Eq.
(6) implies Eqs. (2)-(4). In Eq. (6), the    $S'$ and $E$ systems
are in entangled pure state; for $\rho_S^2 = \rho_S$, the $\rho_S$
state is also pure.  The entanglement is due to the interaction
$H_{S'E}$, eq. (3), i.e. due to the fact that the environment
effectively monitors and purifies the $S'$ system.

As we show next, the state Eq. (6) is in intimate relation to
quantum information localization measured by "locally inaccessible
information (LII)" flow [11], as well as with quantum discord
saturation [14] and quantum decorrelation [15, 16].

\noindent {\it Lemma 1}. The following are mutually equivalent
statements: (i) the system $S+S'+E$ is in the state Eq. (6), (ii)
quantum discord $D^{\leftarrow}(S+S'|E) = \mathcal{S}(E)$ is
saturated (maximal), (iii) there is total decorrelation of the $S$
system from the system $S'$ and (iv) there is quantum information
localization in the $S'+E$ system.

We prove this lemma in a way supporting some intuition about the
zero-discord classicality. The more formal and  more simple proofs
will be provided elsewhere.

\noindent Proof: Bearing in mind (i) is equivalent to (ii) (cf.
Theorem 1 in [16]), the proof can be given by proving  (i) is
equivalent to (iii) and to (iv). That (i) implies (iv) is easy
obtained. The "locally inaccessible information" flow [11],
$\mathcal{L}^{\leftrightarrow} = D^{\leftrightarrow}(S'|S) +
D^{\leftrightarrow}(E|S') + D^{\leftrightarrow}(S|E) =
D^{\leftrightarrow}(E|S')$; there is only information flow in
$S'+E$ system. Now we prove the inverse to this implication. Due
to non-negativity of discord, the above equality for
$\mathcal{L}^{\leftrightarrow} $ directly implies
$D^{\leftrightarrow}(S|S') = 0 = D^{\leftrightarrow}(S| E)$. As we
know $D^{\leftrightarrow}(S'| E) \neq 0$, the condition
$D^{\leftrightarrow}(S|S') = 0 = D^{\leftrightarrow}(S|E)$ can be
satisfied  only by the state Eq. (6); e.g., the alternative
tripartite state, $\sum_i c_i \vert i \rangle_S \vert i
\rangle_{S'} \vert i \rangle_E$, that satisfies
$D^{\leftrightarrow}(S|S') = 0 = D^{\leftrightarrow}(S| E)$, does
not satisfy $D^{\leftrightarrow}(S'| E) \neq 0$. Here (without
loss of generality) we assume the total system $S+S'+E$ is subject
to the Schr\" odinger law, cf. Eq. (4), and that the initial
states of both $S'$ and $E$ are pure--thus the alternative mixed
states are of no interest here. Finally, we prove equivalence of
(i) and (iii). The decorrelation is defined [15, 16] as a
difference of the two total correlations in the initial and the
final state, $I_{initial}(S: S') - I_{final}(S:S')$. For every
initial state, decorrelation is maximal if $I_{final}(S:S') = 0$.
So, we prove that $I_{final}(S: S') = 0$ is equivalent to Eq. (6).
From Eq. (6) it directly follows: $I(S:S') = \mathcal{S}(S) +
\mathcal{S}(S') - \mathcal{S}(S, S') = 0$. The inverse is easily
proved, as from $I(S:S') = 0$ follows $\mathcal{S}(S,S') =
\mathcal{S}(S) + \mathcal{S}(S')$, which, in turn, is fulfilled
only for the product states, Eq. (5). By purifying the product
state, Eq. (5), one obtains the state Eq. (6). This completes the
proof.

\smallskip

The proof of Lemma 1 distinguishes the physical relevance of the
model Eq. (2). Saturation of quantum discord (in $S'+E$) is
equivalent to locking information locally (in $S'+E$), i.e. to
decorrelation of the rest ($S$) of the composite system. So,
Markovian classicality of $S+S'$ coincides with quantumness of
$S'+E$. Of course, external influence on $S'+E$ leads to the loss
of maximum discord. Bearing in mind the result of Ferraro et al
[7], cf. Introduction, Lemma 1 suggests the locking of information
[11], discord saturation [14] and quantum decorrelation [15, 16]
are dynamically feasible only asymptotically.

The model Eqs. (2)-(5) is in accordance with the following logic
of the decoherence theory [19, 24]: only certain degrees of
freedom ($S'$) of a composite system are subject to decoherence.
The remaining degrees of freedom (the $S$ system) can exhibit
quantum mechanical behavior.

On the other hand,  the total system, $S+S'+E$,  is not allowed to
correlate with any outer system, denoted by $W$. This is a direct
consequence of the discord saturation,
$D^{\leftrightarrow}(S+S'|E) = \mathcal{S}(E)$, the point (ii) of
Lemma 1. The discord saturation implies non-correlation of $E$
with $W$ [14]. Furthermore, both the $S$ system and the $S'$
system are uncorrelated with the outer $W$ system, in every
instant in time. This conclusion follows from the very
construction of the model Eq. (2). Namely, the S system is closed,
while interaction of the S' system with the $W$ system would
correlate $E$ and $W$, in contradiction with the saturation of
$D^{\leftarrow}(C|E)$. Of course, isolation of $S+S'+E$ from the
rest of the world, $W$, is physically crude and naive.

\bigskip

{\bf B. Quantum Structures}

\bigskip

The following objection is  in order: for the realistic particles
that mutually interact, one can hardly expect  isolation as
presented by the $S$ system. To answer, we need a switch in
perspective to describing the composite systems.

\smallskip

\noindent {\it Definition 2}. A set of subsystems of a composite
system, $C$, is called a {\it structure} of $C$. Different
structures are mutually related by the proper canonical
transformations (CTs), which provide the different tensor-product
forms for the system's Hilbert space.

\smallskip

The CTs induce a change in both the composite system's
Hilbert-space tensor-product form as well as in the system's
Hamiltonian form. Regarding Figs. 1 and 2, for the Hilbert state
space of the composite system, $\mathcal{H}$, one can write:
$\mathcal{H}_1 \otimes \mathcal{H}_2 = \mathcal{H} = \mathcal{H}_S
\otimes \mathcal{H}_{S'}$. The Hamiltonian, $H$, takes the
different forms for the different structures, $H_1 + H_2 + H_{12}
= H = H_S + H_{S'}$; $H_{12}$ is interaction term, while the
analogous term is absent for $S+S'$ structure.

While the composite system's Hilbert state-space, the Hamiltonian
and quantum state is unique (in every instant in time), the
correlations (for isolated or open system) are {\it not}. This
{\it correlations relativity} formally means [25]: correlations
(quantum or classical, for isolated or open system) are in general
not invariants of the CTs. In other words: the amount of
correlations in {\it instantaneous} state of $C$ is {\it not} a
matter of a composite system itself, {\it but} a matter of the
composite system's structure.

Grouping subsystems (the "coarse graining" the composite system's
structure) is formally a trivial kind of CTs. Entanglement
swapping (see Introduction) is typical of this kind of CTs that
the initially tensor-product form of a state transforms into
entangled form of the state, for the same instant in time.
Regarding the model Eq. (2), for instantaneous state Eq. (6), the
two bipartite structures, $(S+S')+E$ and $S+(S'+E)$, also bear the
different discords, $D^{\leftarrow}(S + S'\vert E) =
\mathcal{S}(E) \neq 0$ and $D^{\leftarrow}(S\vert S'+E) = 0$,
respectively.

Quantum correlations relativity is also implicit e.g. in the
"entanglement renormalization" methods for the finite-dimensional
many-body systems [26] (and the references therein). A specific
decoupling (variables separation) procedure provides a bipartite
structure Eq. (2) for the system of interacting spins. The
original 'microscopic' degrees of freedom are transformed to
introduce a pair of noninteracting systems. Then the ground energy
(pure) state, that bears entanglement for the 'microscopic'
structure, obtains the tensor-product form.

\begin{figure}
 \includegraphics[width=0.5\textwidth]{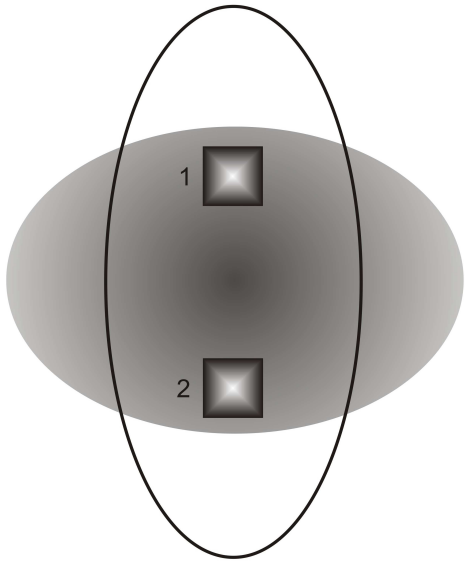}\\
  \caption{Schematic illustration of the $1+2$ structure of the
composite system, $C$, is distinguished by the elliptic line. The
gray area designates the environment $E$ that is not in
interaction with all the $C$'s degrees of freedom. The realistic
particles, $1+2$, degrees of freedom are linked with the degrees
of freedom of $S+S'$ (cf. Fig.1) via the proper canonical
transformations; $1+2 = C = S+S'$. As distinct from the $S+S'$
structure, the  $1+2$ structure may be expected of non-zero
discord.}\label{Fig. 2}
\end{figure}

For the continuous variable (CV) systems, the variables separation
procedure is an open issue in intimate relation to the issue of
(quantum) integrability. To this end, paradigmatic are the
composite system's center-of mass ($CM$) and the "relative
(internal)" degrees of freedom, cf. Figs.1 and 2. Regarding the CV
Gaussian states, the important results in [27] strongly support
the model Eq. (2). As the only zero-discord Gaussian states are
tensor product [27], the only Gaussian states dynamically
supporting Markovian classicality--are of the form Eq. (5). Of
course, for non-Gaussian states the things may generally look
different.

So the physical relevance of the  model $S+S'$, Eq. (2), follows
also from its {\it universal} applicability--just transform the
"original" structure $1+2$ into a structure formally presented by
Eq. (2). The systems $S$ and $S'$, Fig.1, can represent
respectively the "relative (internal)" ($R$) and the
center-of-mass ($CM$) degrees of freedom for the pair $1+2$,
Fig.2, or the original spin-chain and a pair of noninteracting
blocks [26].  The composite systems not describable by the
structure Eq. (2) may be practically deprived of the zero discord
states.

\bigskip

{\bf C. Summary}

\bigskip

In {\it support} of the model, we distinguish: a. the model is in
accordance with the general logic that only a subset of the
open-system's degrees of freedom ($S'$) is subject to decoherence;
b. $S+S'$ resembles the classical-mechanics model-structure in the
general use, $CM+R$, Section III.B; c. regarding the Gaussian
states, the results in [27] strongly support the model.

On the other hand, the model can be considered {\it too crude and
idealized},  as: d.  exact separation of the $S$ system from the
rest in Eq. (2) does not seem very realistic; e. in disagreement
with the general logic of the open system and decoherence theory,
the model does not allow approximate isolation of (i.e. the
information flow from and to) the total system $S+S'+E$.

In conclusion of this section we define a new task that is a
subject of the next section: to search for a variation, i.e.
approximation, of the model in order to avoid the objections 'd'
and 'e', while saving its virtues, the above points 'a-c'.

\bigskip

{\bf IV. APPROXIMATE MARKOVIAN CLASSICALITY}

\bigskip

\noindent {\it Definition 3.}  An open quantum system, $C$,
consisting of two subsystems, $S$ and $S'$, is said to bear {\it
approximate Markovian classicality} if and only if it can be
described by a approximate classical-classical (CC) state in a
sufficiently long time interval.

"Approximate CC state" is a state that can be approximated  by an
CC state, Definition 1. "Sufficiently long time" emphasizes the
time interval for validity of the approximate Markovian
classicality (AMC) is long compared to the time intervals
characteristic for certain physical processes of interest, but
shorter than the open system's relaxation time, if it is defined
for the model. Now we formulate:

\smallskip

\noindent Approximate Classicality Constraint: {\it Two-way
quantum discord is approximately  zero in a sufficiently long time
interval before eventual thermalization of the open system.}

{\it Prima facie}, one could expect that  nonzero discord will
dynamically quickly become non-negligible [7, 20-22]. On the other
hand, having in mind the obstacles emphasized in Section II, it is
not obvious where, and which kind of approximations can be
 made in order to provide AMC.
 Nevertheless, below we emphasize that a model of the quantum
 information locality [13] implements also the approximate
 Markovian classicality.

The dynamic model in [13] is formally a variant of the model Eq.
(2): In a tripartite system, $S, S', E$, the interaction between
$S'$ and $E$ dominates the composite system's dynamics. The $S$
system interacts with $S'$ but not with the environment $E$. For a
special initial state, $\vert p \rangle_{S'}$, the total system's
state can be presented (see Appendix for some details) in the
following simplified form [13, 28]:

\begin{equation}
\vert \Phi^p\rangle_{SS'E} = \vert \psi_p (t) \rangle_S \otimes
\vert p \rangle_{S'} \otimes \vert \phi_p(t) \rangle_E + \vert
O(\epsilon, p; t) \rangle_{S+S'+E}.
\end{equation}

In Eq. (7): $\epsilon \equiv c/C \ll 1$, where $c$ is the strength
of interaction between $S$ and $S'$, and $C$ is the interaction
strength for $S'$ and $E$. The first term in
 Eq. (7) is totally-tensor-product state in the time interval  $\tau \sim \epsilon^{-1}$. In the limit $\epsilon \to 0$, Eq.
(7) becomes a variant of Eq. (6).

 Physically, the state Eq. (7) provides approximate
separation of all subsystems and therefore very small discord for
the open system $C=S+S'$ in the time interval $\tau \sim C/c$.
This is a direct consequence of the following lemma.

\smallskip

\noindent {\it Lemma 2}. Von Neumann entropy of every subsystem in
Eq. (7) is proportional to $\epsilon$, in the time interval $\tau
\sim C/c$.

\noindent Proof: The tripartite system can be decomposed as a
bipartite system by grouping, e.g. $(S+S') + E$. Then the
(normalized) state Eq. (7) takes the form $\sqrt{1 - \epsilon}
\vert \phi \rangle_{SS'} \vert \chi \rangle_E + \sqrt{\epsilon}
\sum_i \sqrt{p_i} \vert i \rangle_{SS'} \vert i \rangle_E$,
$\sum_i p_i = 1$, for every instant in time. For $\rho_{SS'} =
tr_E \vert \Phi\rangle_{SS'E}\langle \Phi \vert$, the $S+S'$
entropy, $\mathcal{S}(S+S') = - (1-\epsilon) \ln (1-\epsilon) -
\sum_i \epsilon p_i \ln (\epsilon p_i) \sim \epsilon (1 - \ln
\epsilon - \sum_i p_i\ln p_i) \le  \kappa \epsilon$, $\kappa
\equiv 1 - \ln \epsilon - \ln p_{max}$, where $p_{max} =
\max\{p_i\}$. As $\mathcal{S}(E) = \mathcal{S}(S+S')$ and the
analogous result follows for the other bipartite decompositions,
$S+(S'+E)$ and $S'+(S+E)$, Lemma 2 is proved.

Now it is straightforward to see the total mutual information in
the $S+S'$ system, $I(S:S') = \mathcal{S}(S) + \mathcal{S}(S') -
\mathcal{S}(S,S')$, is proportional to $\epsilon$.  Due to
nonnegativity of the discord and of the classical correlations
(cf. Section II), it is clear that the discord is also
proportional to $\epsilon$ and is very small.  The only exception
is the case of the maximum entanglement in the small term in
$\vert \Phi^p\rangle_{SS'E}$. i.e. For $\vert O(\epsilon, p; t)
\rangle_{S+S'+E}=\sum\limits_{i=1}^N p_i \vert i \rangle_{S+S'}
\vert i\rangle_E$ and for
 $p_i = N^{-1}, \forall{i}$,
one obtains $\mathcal{S}(S+S') \approx \epsilon \ln N$ [29], when,
in principle, for given $\epsilon$ there may exist $N$ such that
$\epsilon \ln N \sim 1$. The "locally inaccessible information
(LII)" flow [11] is also negligible,
$\mathcal{L}^{\leftrightarrow} = D^{\leftrightarrow}(S'|S) +
D^{\leftrightarrow}(E|S') + D^{\leftrightarrow}(S|E) \propto
\epsilon$. Like in Ref. [21], we can hope that the model Eq. (7)
may provide both the discord and the LII flow are zero for some
{\it practical purposes} (even though not rigorously null), not
only for the already known purposes of combating decoherence [28,
30] and providing identity of micro-particles in a solution [31].

While respecting the points 'a-c', Section III.C, the following
are the virtues of the  model Eq. (7): 1. the model does not
require yet supports Markovian environment $E$; 2. the model is
generally applicable--it equally targets the finite dimensional as
well as the continuous variable systems; 3. the model allows
interaction between $S$ and $S'$, Eq. (2). This way the model
resolves the above point 'd', Section III.C; 4. Due to the
non-zero second term in Eq. (7), the discord $D^{\leftarrow}(C
\vert E)$ is not saturated (cf. Theorem 1 in Ref. [16]). Therefore
the above point 'e', Section III.C, is also resolved: the
environment $E$ is allowed  to correlate with the external system
$W$.

\bigskip

{\bf V. DISCUSSION}

\bigskip

We are not aware of any constraints on discord dynamics for the
non-Markovian open systems.  In principle, every subsystem ($1, 2,
S, S'$) in the model Eq. (2) may be  a composite system itself.
Related multi-partitions of the total system $C$ require separate
analysis to be presented elsewhere.

 Essential for our model of
Markovian classicality are the assumptions on the special initial
state (the tensor-product state) and the completely positive
dynamics of the open system. Regarding the initial state, our
assumption is of general use for
 noninteracting systems--from the hydrogen atom and the "ideal
gas" to the macroscopic bodies typically modelled by their center
of mass and the internal degrees of freedom (the $CM+R$
structure).

The model Eq. (2) is already in use. An ancilla qubit, appearing
in a number of the quantum information protocols and algorithms
[18], is easily recognized as the $S$ system in the model Eq. (2),
which, in turn,  has recently been proposed as a testbed for
investigating non-Markovian dynamics of open systems [32].
"Entanglement renormalization" provides decoupling in a bipartite
structure for a spin-chain [26] thus providing the tensor-product
form Eq. (2) for the ground state of the spin chain. In general,
nonexistence of the model Eq. (2) for a concrete physical system
suggests the composite system is practically deprived of the zero
discord states. To this end, existence of an alternative to the
model Eq. (2) may vary the conclusions. However, bearing in mind
the tight conditions for Markovian classicality, Section II, we
are free to conjecture nonexistence of alternate model that would
{\it rigorously} provide Markovian classicality. Nevertheless,
alternate models of approximate Markovian classicality can be
expected.

The approximate Markovian classicality model, Section IV, suggests
physically there is not ideal Markovian classicality. Worse,
approximate Markovian classicality can last for only a finite time
interval. To this end, the details are case sensitive and
establishing approximate Markovian classicality for some practical
purposes can hardly be formulated in full generality.

Our results have wide-range implications. First, from a
fundamental perspective, they imply that only-classically
correlated states are the matter of the open system's structure.
Bearing in mind Lemma 1, we realize that non-asymptotic
information locality [11], discord saturation [14] and quantum
decorrelation [15, 16] are also the matter of the composite
system's structure. The composite systems not allowing such
structure may be  deprived of the zero discord states. Second,
completely positive dynamics is a necessary condition for the
model Eq. (6), Eq. (7). So, a change in structure of the composite
system removes the conflict [7]  between the completely positive
map and the rarity of zero discord states.  Third, by Definition 1
(Definition 3) the model Eq. (6) (Eq. (7)) is not a model of the
sudden death of discord--discord is zero (or approximately zero)
in a long time interval without the sudden change or sudden death.
Fourth, the model Eq. (6) directly sets  a basis for the task of
"local broadcasting" [33].

A final comment about experimental implications. Very much like
the classical systems, the structure $S+S'$ described by the model
Eq. (6), or Eq. (7), is not capable of performing a useful quantum
information processing. So, instead of experimentally testing
discord (that is not feasible [7] yet), one can try to perform
quantum information processing. The failure of every possible
quantum protocol, e.g. of the discord-based quantum computation
[5, 6] (and the references therein), reveals, at least
approximate, Markovian classicality of the composite system's
structure. Thereby, avoiding Markovian classicality now appears
basic to performing efficient information processing on
bipartitions of the quantum information hardware.

\bigskip

{\bf VI. CONCLUSION}

\bigskip

Complementary to Ferraro et al [7], we construct a model of a
Markovian bipartite system that provides zero two-way discord in a
long time interval (that we name 'Markovian classicality'). The
model is a sufficient condition for Markovian classicality and is
in close relation to the  topics of quantum information locality,
discord saturation and quantum decorrelation. We emphasize
Markovian classicality is not a matter of the open system itself,
but of the open system's structure (decomposition into
parts/subsystems). Bearing this in mind, the model is of general
interest and is not a model of quantum discord sudden death yet.
We finally conjecture about the absence of alternate model, which
would rigorously meet the criteria for Markovian classicality.

\bigskip

{\bf ACKNOWLEDGEMENTS}

\bigskip

\noindent We benefited much from discussions with  P. J. Coles, S.
Alipour and C. A. Rodr\' iguez-Rosario,. The work on this paper is
financially supported by Ministry of Science Serbia under contract
no 171028.

\bigskip

[1] H. Ollivier, W. H. Zurek, Phys. Rev. Lett. {\bf 88}, 017901
(2001)

[2] L. Henderson, V. Vedral,  J.  Phys. A: Math. Gen. {\bf 34},
6899 (2001)

[3] B. Daki\' c, V. Vedral, \v C. Brukner, Phys. Rev. Lett. {\bf
105}, 190502 (2010)

[4] S. Luo, Phys. Rev. A 77 (2), 022301 (2008)

[5] K. Modi, A. Brodutch, H. Cable, T. Paterek, V. Vedral, 2011,
arXiv:1112.6238v1 [quant-ph]

[6] L. C. C\' eleri, J. Maziero, R. M. Serra, Int. J. Qu. Inform.
{\bf 9}, 1837 (2011); J.-S. Xu, C.-F. Li, 2012, arXiv:1205.0871v1
[quant-ph]

[7] A. Ferraro, L. Aolita, D. Cavalcanti, F. M. Cucchietti, A.
Acin, Phys. Rev. A {\bf 81}, 052318 (2010)

[8] M. Arsenijevi\' c, J. Jekni\' c-Dugi\' c, M. Dugi\' c, 2012,
arXiv:1201.4975v3 [quant-ph]

[9] P. J. Coles, Phys. Rev. A {\bf 85}, 042103 (2012)

[10] A. Ferraro, M. G. A. Paris, 2012, arXiv:1203.2661v1
[quant-ph]

[11] F. F. Fanchini, L. K. Castelano, M. F. Cornelio, M. C. de
Oliveira, New J. Phys. {\bf 14}, 013027 (2012); F. F. Fanchini, M.
F. Cornelio, M. C. de Oliveira, A. O. Caldeira, Phys. Rev. A {\bf
84}, 012313 (2011)

[12] D. Beckman, D. Gottesmann, M. A. Nielsen, J. Preskill, Phys.
Rev. A {\bf 64}, 052309 (2001); B. Schumacher, M. D. Westmoreland,
Qu. Inf. Processing {\bf 4}, 13 (2005)

[13] M. Dugi\' c, J. Jekni\' c-Dugi\' c, Chin. Phys. Lett.  {\bf
26}, 090306 (2009)

[14] Z. Xi, X.-M. Lu, X. Wang, Y. Li, Phys. Rev. A {\bf 85},
032109 (2012)

[15] S. Luo, S. Fu, N. Li, Phys. Rev. A {\bf 82}, 052122 (2010)

[16] L. Zhang, J. Wu, J. Phys. A: Math. Theor. {\bf 45},
025301(2012)

[17] C. H. Bennett, G. Brassard, C. Crepeau, R. Jozsa, A. Peres,
W. K. Wootters, "Teleporting an Unknown Quantum State via Dual
Classical and Einstein-Podolsky-Rosen Channels" Phys. Rev. Lett.
70, 1895-1899 (1993)

[18] M. A. Nielsen, I. L. Chuang, "Quantum Computation and Quantum
Information" (Cambridge Univ. Press, Cambridge, 2000).

[19] H. P. Breuer, F.  Petruccione, "The Theory of Open Quantum
Systems" (Clarendon Press, Oxford, 2002)

[20] S. Campbell, T. J. G. Apollaro, C. Di Franco, L. Banchi, A.
Cuccoli, R. Vaia, F. Plastina, M. Paternostro, Phys. Rev. A {\bf
84}, 052316 (2011); A. Streltsov, H. Kampermann, D. Bruss, Phys.
Rev. Lett. {\bf 107}, 170502 (2011); F. Ciccarello, V.
Giovannetti, Phys. Rev. A {\bf 85}, 010102(R) (2012); X. Hu, Y.
Gu, Q. Gong, G. Guo, Phys. Rev.  A {\bf 84}, 022113 (2011); M.
Gessner, E.-M. Laine, H.-P. Breuer, J., Piilo, 2012,
arXiv:1202.1959v1 [quant-ph]

[21] F. Ciccarello, V. Giovannetti, Phys. Rev. A {\bf 85}, 022108
(2012)

[22] S. Tesfa, Optics Communications {\bf 285}, 830 (2012)

[23] C. A. Rodr\' iguez-Rosario, G. Kimura, H. Imai, A.
Aspuru-Guzik, Phys. Rev. Lett. {\bf 106}, 050403 (2011)

[24] D. Giulini, E. Joos, C. Kiefer, J. Kupsch, I.-O. Stamatescu
and H. D. Zeh, {\it Decoherence and the Appearance of a Classical
World in Quantum Theory} (Berlin: Springer,1996)

[25] M. Dugi\' c, M. Arsenijevi\' c, J. Jekni\' c-Dugi\' c, 2011,
2011, arXiv:1112.5797v3 [quantph]

[26] G. Evenbly, G. Vidal, 2012, arXiv:1205.0639v1 [quant-ph]

[27] G. Adesso, A. Datta, Phys. Rev. Lett. {\bf 105}, 030501
(2010); P. Giorda, M. G. A. Paris, Phys. Rev. Lett. {\bf 105},
020503 (2010)

[28] M. Dugi\' c, Quantum Computers and Computing {\bf 1}, 102
(2000)

[29] N. Linden, S. Popescu, J. A. Smolin, Phys. Rev. Lett. {\bf
97}, 100502 (2006).

[30] J. Busch, A. Beige, J. Phys.: Conf. Ser. {\bf 254}, 012009
(2010)

[31] M. Dugi\' c, Europhys. Lett. {\bf 60}, 7 (2002)

[32] \' A Rivas, S. F. Huelga, M. B. Plenio, Phys. Rev. Lett. {\bf
105}, 050403 (2010); S. Alipour, A. Mani, A. T. Rezakhani, 2012,
arXiv:1203.2347v2 [quant-ph]

[33] M. Piani, P. Horodecki, and R. Horodecki, Phys. Rev. Lett.
{\bf 100}, 090502 (2008).

\bigskip

{\bf Appendix}

\bigskip

Originally, the DISD model is developed for the purposes of
combating decoherence in quantum computation hardware.

It's a tripartite system of interest, $S+S'+E$, defined by the
Hamiltonian:

\begin{equation}
H = H_S + H_{S'} + H_E + H_{SS'} + H_{S'E},
\end{equation}

\noindent where the double subscript denotes interactions. While
not assuming anything about any of the subsystems, $S$, $S'$ and
$E$, the model assumes the interaction $H_{S'E}$ dominates the
total system's dynamics and the system $S'$ is in the initial
state $\vert p \rangle_{S'}$ satisfying the "robustness"
condition, $H_{S'E} \vert p \rangle_{S'} \vert \chi \rangle_E =
\vert p \rangle_{S'} \vert \chi_p \rangle_E$; the strength of
$H_{S'E}$ is denoted by $C$ while the strength of interaction
$H_{SS'}$ is denoted by $c$.

Then the use of the standard perturbation procedure for the
(normalized) initial state $\vert \Psi^p \rangle_{SS'E} = \sum_k
C_k \vert k\rangle_S \otimes \vert p \rangle_{S'} \otimes \sum_j
\beta_j \vert j \rangle_E$, one obtains the {\it exact}
total-system's state:

\begin{eqnarray}
&\nonumber& \vert \Psi^p(t) \rangle_{SS'E} = \left( \sum_k C_k(t)
\exp (-\imath t\lambda^1_{kp}/\hbar) \vert k \rangle_S \right)
\otimes \exp(- \imath t \lambda / \hbar) \vert p \rangle_{S'}
\\&& \otimes
\left(\sum_{j} \beta'_j(t) \exp(- \imath t  \lambda^{kpj}/\hbar)
\vert j \rangle_E\right) + \vert O(\epsilon,t)\rangle_{SS'E}.
\end{eqnarray}

In Eq. (14): $C_k(t) \equiv C_k \exp(-\imath t _S\langle k\vert
H_S \vert k\rangle_S/\hbar)$, $\lambda \equiv _{S'}\langle p \vert
H_{S'} \vert p \rangle_{S'}$, $\beta'_j(t) \equiv \beta_j
\exp(-\imath t (C\kappa_{kj} + _E\langle j \vert H_E \vert j
\rangle_E))$. $\kappa_{pj}$ represents an eigenvalue of $H_{S'E}$,
$\lambda^1_{pk} = _{SS'}\langle pk \vert H_{SS'} \vert
pk\rangle_{SS'}$ is the first-order correction and $\lambda^{kpj}$
is of the order of the second-order correction to the eigenvalues
of $H_{S'E}$, while $\epsilon \sim c/C$. Due to $\lambda^{kpj}
\sim c/C$, after a time interval $\tau' > C/c$ , the induced
correlations of $S$ and $E$ become non-negligible.

\pagebreak

{\bf SUPPLEMENTAL MATERIAL}

\bigskip

 The Supplemental Material provides proofs of
our results and justification for various remarks that we made in
the main manuscript. The Supplemental Material does not attempt to
follow the chronology of the main manuscript. Rather, it is
devoted to providing description of the two structures, cf. Figs.
1 and 2, in the main manuscript. It consists of three parts. The
first part is devoted to calculating discord with an emphasis on
the two structures.  The second part exhibits subtlety of the
recipe for constructing the structure Eq. (2). The third part
discusses importance of the collective variables of the
center-of-mass and the internal degrees of freedom and is in
support of the Conjecture made in Section V.

\bigskip

{\bf Calculating discord for different structures}

\bigskip

We are concerned with the two possible decompositions of the total
system, $S+S'+E$. We calculate the states, their entropies and the
related quantum discords for both structures separately. For
simplicity, by $\mathcal{S}(A)$ we denote von Neumann entropy of
the $A$ system's state, $\rho_A$, $\mathcal{S}(A) \equiv
\mathcal{S}(\rho_A)$.

Quantum state Eq. (6) can be written as:

\begin{equation}
\rho \equiv \rho_S \otimes \sum_{i,j} c_i c^{\ast}_j \vert i
\rangle_{S'} \langle j\vert \otimes \vert i\rangle_{E} \langle
j\vert .
\end{equation}

\noindent {\bf A.} Structure $(S+S')+E$, denoted by $C+E$, is a
bipartite system and the subsystems' states are:

\begin{eqnarray}
&\nonumber& \rho_{S+S'} = tr_E \rho = \rho_S \otimes \sum_i \vert
c_i \vert^2 \vert i\rangle_S \langle i\vert; \quad \rho_E =
tr_{S+S'} \rho = \sum_i \vert c_i \vert^2 \vert i\rangle_E \langle
i\vert \\&& \quad \rho_{S'} = tr_{S+E} \rho = tr_S \rho_{S+S'} =
\sum_i \vert c_i \vert^2 \vert i\rangle_{S'} \langle i\vert.
\end{eqnarray}

Then von Neumann entropies, $\mathcal{S}(S,S') = \mathcal{S}(S) +
\mathcal{S}(S')$ and $\mathcal{S}(S') = \mathcal{S}(E) = \sum_i
\vert c_i\vert^2 \log \vert c_i \vert^2$. From Eq. (11) we
directly obtain $\rho_{S+S'}\vert_i \equiv I_{S+S'} \otimes \vert
i\rangle_E\langle i \vert \rho  I_{S+S'} \otimes \vert
i\rangle_E\langle i \vert = \rho_S \otimes \vert i
\rangle_{S'}\langle i \vert$, assuming the measurement in the
$\{\vert i \rangle_E\}$ basis is performed, as well as
$\mathcal{S}(\rho_{S+S'}\vert_i) = \mathcal{S}(S) +
\mathcal{S}(S') = \mathcal{S}(S)$. For the total system's entropy,
Eq. (12) gives for, in general, mixed state $\rho_S =
\sum_{\alpha} \omega_{\alpha} \vert \alpha \rangle_S\langle \alpha
\vert$:

\begin{eqnarray}
&\nonumber& \mathcal{S}(S,S',E) = - tr_{S+S'+E} \rho \ln \rho =  -
tr_{S+S'+E}\sum_{i,j,\alpha,k,l\beta} c_i c^{\ast}_j
\omega_{\alpha} [\omega_{\beta}\ln c_k c^{\ast}_l  \\&& + c_k
c^{\ast}_l \ln \omega_{\beta}] \vert i\alpha\rangle_{S+S'}\langle
j\alpha\vert \otimes \vert i\rangle_E\langle j\vert = -
\sum_{\alpha} \omega_{\alpha} \ln \omega_{\alpha} =
\mathcal{S}(S),
\end{eqnarray}

\noindent where we made use of $tr_{S+S'+E} = tr_S tr_{S'} tr_E$
and the basis independence of the tracing out operation, while
$\vert i \alpha \rangle_{S+S'} \equiv \vert i\rangle_S \vert
\alpha \rangle_{S'}$.

Then the total correlations, $I(S,S':E)  = \mathcal{S}(S,S') +
\mathcal{S}(E) - \mathcal{S}(S,S',E)$ and the classical
correlations $J^{\leftarrow} (S,S'\vert E) = \mathcal{S}(S,S') -
\inf_{\{\Pi_{Ei}\}} \sum_i \vert c_i \vert^2
\mathcal{S}(\rho_{S+S'}\vert_{\Pi_{Ei}})$. With the use of the
above calculated entropies, we obtain for the one-way discord (cf.
Section II):

\begin{eqnarray}
&\nonumber& D^{\leftarrow}(S,S'\vert E) = \mathcal{S}(S,S') +
\mathcal{S}(E) - \mathcal{S}(S,S',E) - \mathcal{S}(S,S')
\\&& + \sum_i \vert c_i\vert^2 \mathcal{S}(\rho_{S+S'}\vert_{i}) =
\mathcal{S}(E) - \mathcal{S}(S,S',E) + \sum_i \vert c_i \vert^2
\mathcal{S}(\rho_{S+S'}\vert_i) = \mathcal{S}(E),
\end{eqnarray}

\noindent that is the discord saturation discussed in Section
III.A.

\smallskip

\noindent {\bf B.} Alternative structure $S+(S'+E)$ is more easy
to handle. Then Eq. (6) is of direct use and the results follow:

\begin{eqnarray}
&\nonumber& \rho_S = tr_{S'+E} \rho; \quad \rho_{S'+E} = tr_S \rho
= \vert \Psi\rangle_{S'+E}\langle \Psi \vert \\&&
 \rho_{S'} = \sum_i
\vert c_i \vert^2 \vert i \rangle_{S'}\langle i \vert; \rho_E =
\sum_i \vert c_i \vert^2 \vert i \rangle_E\langle i \vert.
\end{eqnarray}

From Eq. (15) it easily follows: $\mathcal{S}(S,S',E) =
\mathcal{S}(S)  + \mathcal{S}(S',E)$, while $\mathcal{S}(S',E) =
0$ and $\mathcal{S}(S') = \mathcal{S}(E) = - \sum_i \vert c_i
\vert^2 \ln \vert c_i \vert^2$.

Then the total correlations, $I(S:S',E) = \mathcal{S}(S) +
\mathcal{S}(S',E) - \mathcal{S}(S,S',E) = 0$. As both quantum
discord and the classical correlations are non-negative, the
one-way discord $D^{\leftarrow}(S\vert S'+E) = 0$ as well as
$D^{\rightarrow}(S\vert S'+E) = 0$--the state Eq. (6) is a CC
state, Definition 1 in the main manuscript. The third structure,
$(S+E) + S'$, can be alternatively managed with the conclusion
that $D^{\leftarrow}(S+E\vert S') = \mathcal{S}(S') =
\mathcal{S}(E) \neq 0$.

It is interesting that even the trivial change of structure, by
simply grouping the constituent subsystems, $S$, $S'$ and $E$,
exhibits the general notion of quantum correlations relativity
[1]: quantum discord is a matter of structure, and is here zero
only for the $S+(S'+E)$ bipartite structure. The different
structures reveal the different facets of the total system.
Further examples are given in the remainder of the Supplemental
Material.

Finally, we show that the tensor product state Eq. (5), {\it
considered as a $P$-classical state} [2], satisfies the
$C$-criterion for classicality. The generic $P$-classical state
Eq. (1) in Ref. [2] reduces to the tensor-product state
(considered in Section III.B) for the separability condition [in
their notation] $P(\alpha, \beta) = P(\alpha) P(\beta)$. For this
choice, one obtains for the states considered, Eq. (6) in Ref.
[2]: $\rho_A \equiv tr_B \rho_A \otimes \rho_B = \rho_A$, and
(after normalization) $\rho_{\circ} \equiv tr_B \rho_A \otimes
\rho_B \vert 0 \rangle_B\langle 0 \vert = \rho_A$. So, one obtains
$[\rho_A, \rho_{\circ}] = 0$, that is the criterion for the
$C$-classicality, which, in turn, is already well-known. Our proof
is given in terms of $P$-classical states in order to match the
considerations in Ref. [2].

\bigskip

{\bf Constructing the Markovian classicality model}

\bigskip

Consider a composite system $C$ consisting of $N$ physical
particles, $1,2 , 3, ... , N$. Then the set $\mathcal{C}_1 =
\{1,2,3,...,N\}$ is a structure describing $C$ as a multiparticle
system. The set of the $C$'s degrees of freedom, $\{x_{i\alpha},
i=1,2,...N\}$, can be transformed to provide a new structure of
$C$; the index $\alpha$ enumerates the individual particles
degrees of freedom. E.g. by grouping the particles into two sets
described by their degrees of freedom, $A = \{x_{i\alpha}, i =
1,2,...M\}$ and $B = \{x_{i\alpha}, i=M+1, M+2,...,N\}$, we obtain
a bipartite structure of $C$, presented formally as $\mathcal{C}_2
= A + B$. This grouping the particles is kind of formally trivial
canonical transformations (CTs). Formally nontrivial kind of the
CTs assume non-local symplectic transformations that introduce the
new degrees of freedom, $\{\xi_{p\beta}, p=1,2,...,N\}$. To this
end, paradigmatic are the CTs introducing the $C$'s center-of-mass
($CM$) and the relative positions ($R$) degrees of freedom. Then
$C$ can be described by another bipartite structure,
$\mathcal{C}_3 = CM+R$.

Below we briefly discuss the task of constructing the structure
$S+S'$.

In general, the proper canonical transformations convert a
bipartite-system's 'fundamental' structure $1+2$ (Fig. 2) into the
 structure $S+S'$ (Fig. 1). While the system's
Hamiltonian, $H_C$, is unique, it obtains different forms for the
different structures: $H_1 + H_2 +H_{12} = H_C = H_S + H_{S'}$.

The  structure $S+S'$ follows from the variables separation for
the original $1+2$ structure. This is closely related to the
general mathematical topic of integrability of quantum mechanical
models. Regarding the "mixed" states (described by the density
matrix), this is an instance of the task of the Quantum
Separability Problem (QUSEP). QUSEP is investigated in the
literature for the finite-dimensional composite systems (see e.g.
Gharibian [3] and references therein) and is computationally a
"strongly NP-Hard" problem [3].

On the other hand, separation of variables is not much more easier
even for the pure states. The task is to obtain the equality

\begin{equation}
 \sum_i c_i \vert i \rangle_1 \vert i \rangle_2 =  \vert \Psi \rangle= \vert \phi\rangle_S\vert \chi\rangle_{S'}
\end{equation}

\noindent for an instantaneous state, $\vert \Psi \rangle$, of the
composite system $C$.

Physically, Eq. (16) assumes there are interactions and therefore
entanglement for the 'fundamental' structure $1+2$ while mutually
noninteracting systems $S$ and $S'$ are described by a
tensor-product state. We believe these easily formulated tasks are
largely intact in the present quantum theory.

\bigskip

{\bf In support of the Conjecture}

\bigskip

Most of the classical physics deals with the collective variables
of the macroscopic bodies, $CM$ and $R$. "Classicality" of the
macroscopic bodies is tacitly assumed for {\it this} kind of
structure of the classical-physics systems. So, the $S+S'$
structure, Section II, naturally resembles the macroscopic-systems
structure $CM+R$.

In addition, we want to emphasize that the model Eqs. (2)-(5)
reflects the general experience with atoms and molecules [4].
Their "relative positions" degrees of freedom are monitored by the
quantum vacuum fluctuations [5] (and the references therein),
while the center-of-mass ($CM$) degrees of freedom are typically
supposed both decoupled from $R$ as well as possibly subject to
the different kinds of the environment (e.g. to the harmonic bath
in quantum Brownian motion [5], and the references therein). In
order to describe this, introducing the $S$'s environment,
$\mathcal{V}$, into the model Eq. (2) is straightforward: as long
as the two environments, $E$ and $\mathcal{V}$, are decoupled from
each other, nothing changes in our considerations, except the $S$
system  is now described by a proper master equation providing
$\rho_S(t)\neq \rho^2_S(t)$. Finally, as distinguished above, the
tensor-product state Eq. (5) satisfies both $P$- and $C$-criteria
[2] for classicality.

\bigskip

[1] M. Dugi\' c, M. Arsenijevi\' c, J. Jekni\' c-Dugi\' c,
arXiv:1112.5797v3 [quant-ph]

[2] A. Ferraro, M. G. A. Paris, arXiv:1203.2661v1 [quant-ph]

[3] S. Gharibian, Quantum Inf. and Comp. {\bf 10}, 343 (2010)

[4] G. Fraser, Ed., "The New Physics for the twenty-first century"
(Cambridge University Press, Cambridge, 2006)

[5] H. P. Breuer, F.  Petruccione, "The Theory of Open Quantum
Systems" (Clarendon Press, Oxford, 2002)

\end{document}